\documentclass[12pt]{article}

\usepackage{amsmath}
\usepackage{amssymb}
\usepackage{epsfig}
\usepackage{rotate}

\newcommand{\capdef}{}
\newcommand{\mycaption}[2][\capdef]{\renewcommand{\capdef}{#2}%
        \caption[#1]{{\itshape #2}}} 
\makeatletter
\renewcommand{\fnum@table}{\textbf{\tablename~\thetable}}
\renewcommand{\fnum@figure}{\textbf{\figurename~\thefigure}}
\makeatother

\newcounter{myenumi}

\renewcommand{\themyenumi}{\roman{myenumi}}
{\end{list}}

\newlength{\myem}
\newcounter{mysubequation}[equation]

\makeatletter
\renewcommand{\section}{\@startsection{section}{1}{0em}{-\baselineskip}%
{\baselineskip}{\normalfont\large\bfseries}}
\renewcommand{\subsection}%
{\@startsection{subsection}{2}{0em}{-0.7\baselineskip}%
{0.7\baselineskip}{\normalfont\bfseries}}
\makeatother

\newcommand{\bea}{\begin{eqnarray*}}
\newcommand{\eea}{\end{eqnarray*}}

\newcommand{\GeV}{\,\mathrm{GeV}}

\newcommand{\eV}{\,\mathrm{eV}}

\newcommand{\reu}{{\nu_e\rightarrow\nu_\mu}}

\newcommand{\ruu}{{\nu_\mu\rightarrow\nu_\mu}}

\newcommand{\dm}[1]{{\Delta m^2_{#1}}}

\newcommand{\bc}{\begin{center}}
\newcommand{\ec}{\end{center}}
\newcommand{\Pkt}{$\bullet~$}

\begin{document}

%%%%%%%%%%%%%%%%%%%%%%%%%%%%%%%%%%%%%%%%%%%%%%%%%%%%%%%%%%%%%%%%%%%%%

% the footnote symbols are only redefined for the title page !
\renewcommand{\thefootnote}{\alph{footnote}}

\ \vspace*{-3.cm}
\begin{flushright}
  {\hfill TUM--HEP--383/00}\\
  {MPI--PhT/2000-29}\\
  {\ }
\end{flushright}

\vspace*{0.2cm}

\renewcommand{\thefootnote}{\fnsymbol{footnote}}
\setcounter{footnote}{-1}

{\begin{center}
{\Large\bf A Comparison of the Physics Potential of Future Long Baseline Neutrino
Oscillation Experiments$^*$\footnote{\hspace*{-1.6mm}$^*$Work supported by 
        "Sonderforschungsbereich 375 f\"ur Astro-Teilchenphysik" der 
        Deutschen Forschungsgemeinschaft.}}
\end{center}}
\renewcommand{\thefootnote}{\alph{footnote}}

\vspace*{.3cm}
{\begin{center} {\large{\sc
                K.~Dick\footnote[1]{\makebox[1.cm]{Email:}
                Karin.Dick@physik.tu-muenchen.de}\footnotemark[2],~
                M.~Freund\footnote[3]{\makebox[1.cm]{Email:}
                Martin.Freund@physik.tu-muenchen.de},~  
                P.~Huber\footnote[4]{\makebox[1.cm]{Email:}
                Patrick.Huber@physik.tu--muenchen.de}~and~
                M.~Lindner\footnote[5]{\makebox[1.cm]{Email:}
                Manfred.Lindner@physik.tu--muenchen.de}~
                }}
\end{center}}
\vspace*{0cm}
{\it 
\begin{center}  
        
\footnotemark[1]${}^,$\footnotemark[3]${}^,$\footnotemark[4]${}^,$\footnotemark[5]%  
       Theoretische Physik, Physik Department, 
       Technische Universit\"at M\"unchen,\\
       James--Franck--Strasse, D--85748 Garching, Germany

\footnotemark[2]% 
       Max-Planck-Institut f\"ur Physik, Postfach 401212, D--80805 M\"unchen, Germany 

\end{center} }

\vspace*{0.5cm}

\begin{quote}
\bc
{\Large \bf Abstract}
\ec
\vspace*{0.2cm}

{ 
We compare the generic physics potential of various combinations of 
conventional Wide Band or Neutrino Factory Beams with different
detectors to determine several oscillation parameters in long 
baseline experiments. For each combination of beam and detector 
we show the precision which can be obtained for the leading 
oscillation parameters $\dm{31}$ and $\sin^2 2\theta_{23}$. 
Furthermore we show the sensitivity to $\sin^2 2\theta_{13}$ 
and the range in $\sin^2 2\theta_{13}$ for which the sign 
of $\dm{31}$ can be extracted via matter effects. The results 
suggest that existing conventional Wide Band Beam and detector 
technology can be used to considerably improve the precision of 
neutrino properties until a neutrino factory will be built.
}

\end{quote}

\vspace*{0.2cm}

% \newpage

\renewcommand{\thefootnote}{\arabic{footnote}}
\setcounter{footnote}{0}

%%%%%%%%%%%%%%%%%%%%%%%%%%%%%%%%%%%%%%%%%%%%%%%%%%%%%%%%%%%%%%%%%%%%%

Recent studies of neutrino factories as powerful neutrino sources 
have shown that precision measurements of neutrino masses and mixings 
are possible in the future \cite{NUFACT}. The development of a neutrino 
factory includes however challenging technological issues and the 
development will certainly take some time. It is therefore of 
interest what can be achieved meanwhile in comparison by using 
conventional neutrino beams. Such improved conventional experiments 
will lead to a better knowledge of neutrino parameters, which might 
be considered as a useful step to simply bridge the time until a 
neutrino factory will be operating. We will however show in this 
paper that conventional setups may be able to compete in some 
aspects with the long baseline neutrino oscillation program 
of neutrino factories. The results of such intermediate experiments 
would also be important since they affect the optimal strategies for 
neutrino factories. This applies especially to the measurement of 
CP-violation at neutrino factories, where further advance information 
on the mixing parameters (especially $\theta_{13}$) would be very
important \cite{CP}. We perform therefore in this paper a global 
comparison of the generic physics potential of several experimental 
setups and we present the resulting precision of the leading oscillation 
parameters ($\Delta m^2_{31}$ and $\theta_{23}$) as well as the 
sensitivity to the sub-leading parameters ($\theta_{13}$ and the 
sign of $\Delta m^2_{31}$). Our analysis includes energy thresholds 
and resolutions and we show the dependence on these parameters. 
Details of the formalism which is used to derive the current results 
can be found in earlier publications \cite{DFHL,FHL,FLPR}.

The experimental setups considered in this study include two different 
beam types, namely

\hspace*{0.5cm}\Pkt conventional Wide Band Beams \\
\hspace*{0.5cm}\Pkt neutrino factory beams

which are assumed to point at three different types of detectors, specifically

\hspace*{0.5cm}\Pkt magnetized iron detectors\\
\hspace*{0.5cm}\Pkt large water or ice Cherenkov detectors (``neutrino-telescopes'') \\
\hspace*{0.5cm}\Pkt ring imaging water Cherenkov detectors

located at large distance on the surface of the earth. We consider thus
very long baselines up to 11200~km and we include therefore the MSW 
matter effects \cite{MSW} with the full earth density profile \cite{STACEY}. 
We study only the detection of muons which are produced by muon neutrinos, 
since electrons and taus are in general experimentally more difficult. Adding 
electron and/or tau channels would of course affect the analysis. The 
generic physics potential of the detectors depends for the considered 
measurements then in general primarily on three parameters, namely the 
energy threshold for muon detection, the energy resolution and of course 
the detector mass. To characterize typical detector concepts via these 
three parameters is of course a rather simplified approximation of a real 
detector, but it permits a simple and effective evaluation and comparison 
of the generic physics potential of very different detector types on a 
common basis. 

The dependence of the analysis on the detector mass is trivial, but 
important, since it determines the event rates and thus the statistical
limitations. The energy thresholds are usually required to lie as low 
as possible, typically at a few GeV, to cover most of the neutrino beam 
and to measure the oscillation parameters best.
In this paper we approximate that both threshold energy and energy
resolution refer to neutrino energies, not muon energies.  For the
energy resolution we use the conservative model of a constant resolution,
where the half width of the smearing function increases linearly with
the energy: $\sigma_E = c \times E$. We refer to the constant $c$ here
as energy resolution in percent.  The results which are obtained in
this way show thus the generic physics potential of a certain
beam--detector combination including beam properties, cross sections
and essential detector features.  We do not include systematic
limitations and backgrounds which are different for each
setup\footnote{Note however that intrinsic backgrounds due to the beam
  are very similar for all detectors.}.  The results for the generic
physics potential, which will be very impressive, are thus in some
cases expected to be reduced in the actual experiment by some
systematics or background issue.

We discuss now in more detail the assumptions about beams and detectors
of our study and the used parameter values. Some of these parameters 
have simple scaling laws so that other values can be easily obtained. 
The dependence on the less trivial parameters (like detector resolution) 
will be shown in our plots. 

The following neutrino beams are considered:
Conventional Wide Band Beams are produced by shooting high energy protons 
with a few hundred GeV onto a massive target \cite{CWBB}. The produced pions 
and kaons are collected via a magnetic lens system and decay after a few 
hundred meter. We assume here that these beams can be pointed anywhere 
on Earth. The resulting beam consists mainly of $\nu_{\mu}$ with 
admixtures of other flavors at the percent level. Reversing the lens 
current produces in principle a beam consisting mainly of $\bar\nu_{\mu}$,
however the systematics of the two beams is different. Examples for 
this type of neutrino beams are the K2K beam, the NUMI beam and the CNGS beam. 
We use in our calculations
as prototype the CNGS spectrum and flux for 
$4.5\cdot10^{19}\thinspace\textrm{pot}$. 
There exist proposals to upgrade this technique for future long baseline 
experiments, these improvements can be considered by properly rescaling 
our results.

Neutrino Factory Beams produced from muon decays in storage rings are
highly collimated and very intense.The beam systematics are well
understood so that high precision measurements of oscillation
parameters and masses, as well as a dedicated test of matter effects,
should be possible. We include in this work a Neutrino Factory beam
from $10^{20}$ useful muon decays at an energy of $50~\GeV$.

Magnetized iron detectors are widely considered to be the detection 
system of choice for neutrino factory long baseline oscillation experiments 
\cite{NUFACT}. Very good charge identification capabilities (CID) are 
required to separate muons from antimuons in order to separate the $\reu$ 
appearance oscillation channel from the $\ruu$ disappearance channel. 
If sufficient CID can not be achieved\footnote{This issue is still open
and one can find claims ranging from $10^{-6}$ up to $10^{-2}$.}, then 
appearance channel measurements are spoiled and the analysis of the 
disappearance channels becomes important \cite{FHL}. To take this issue 
into account, we consider here the two extreme cases: Perfect CID and 
completely missing CID. For this type of detector we use an energy threshold 
of 4~GeV as proposed also in the neutrino factory design studies. Furthermore
we use an energy resolution of $10\%$ and a total mass of $10 \mathrm{ kt}$ 
\cite{MINOS,MONOLITH}.

Large water or ice Cherenkov detectors (known also as ``neutrino
telescopes'') \cite{NUTELESCOPES}, which consist of large arrays of
photomultipliers (PMTs) placed in antarctic ice or sea water, were
until recently not considered for very long baseline neutrino
experiments. This is connected to the fact that these detectors are
usually thought of as having a rather high energy threshold. This high
threshold arises for the reconstruction of cosmic events and is a
consequence of the distances in the PMT multiplier arrays and the
minimum number of PMT hits for track reconstruction (i.e. the
direction of the event).  The threshold is also connected to the
reduction of the background.  It was however recently pointed out,
that a neutrino oscillation experiment with high event rate and known
source position, can have a much lower threshold \cite{DFHL}.
Reconstructing the event direction is in this case in principle not
necessary and one can therefore trigger on fewer PMTs leading to a
lower threshold for this mode of operation. Using the beam pulse
timing information and rough muon direction information it is
furthermore possible to reduce the background and hence also the
energy threshold significantly.  Furthermore we will show in this paper
that some measurements are rather insensitive to this threshold. For
our study we use as prototype detector IceCube with an effective mass
of $100 \mathrm{ Mt}$ and an energy resolution of $50\%$. The energy
threshold should lie somewhere between an optimistic value of 5~GeV
and a conservative value of 30~GeV (which could be easily achieved by
the IceCube project) and we use in our study a medium value of 15~GeV
\cite{DFHL}.

As a last type of detector we  considered a megaton ring imaging Cherenkov 
detector like in the AQUA-RICH proposal. Due to its dense optical 
sensor array and imaging technique, such a detector would have an 
extraordinary small energy threshold. Furthermore with its $1\mathrm{Mt}$
detector mass, AQUA-RICH lies between the above presented detectors. 
We use for such a detector as characterizing parameters an energy threshold 
of 1~GeV, an energy resolution of $7\%$ and an overall mass of 
$1\mathrm{Mt}$ \cite{AQUARICH}.

\begin{table}[htb!]
\label{tab:parameters}
\begin{center}
\begin{tabular}{|c||c|c|c|} \hline
                & {\bf Magnetized}  & {\bf Water/Ice}        & {\bf Megaton Ring}   \\
                &{\bf Iron Detector}&{\bf Cherenkov Detector}&{\bf Imaging Detector}\\ 
\hline\hline
{\bf Threshold} & $4~\mathrm{GeV}$  & $15~\GeV$              & $1~\GeV$             \\
\hline
{\bf Resolution}& $10\%$            & $50\%$                 & $7\%$                \\ 
\hline
{\bf Mass}      & $10~\mathrm{kt}$  & $100~\mathrm{Mt}$      & $1~\mathrm{Mt}$      \\
\hline
{\bf CID}       & yes               & no                     & no                   \\
\hline
{\bf Examples}  & MINOS, MONOLITH   & AMANDA, IceCube,       & AQUA-RICH            \\
                &                   & ANTARES, NESTOR        &                      \\
 \hline
\end{tabular}
\end{center}
\caption{Typical detector parameters.}
\end{table}

%%%%%%%%%%%%%%%%%%%%%%%%%%%%%%%%%%%%%%%%%%%%%%%%%%%%%%%%%%%%%%%%%%%%%

The results which we will present below are obtained in a standard
three neutrino framework in matter. We use for the atmospheric mass
splitting the current best fit, i.e. $\dm{31}=\dm{32}=\dm{} = 3.2\cdot
10^{-3}\thinspace\eV^2$ and we use maximal atmospheric mixing, $\sin^2
2\theta_{23}=1$ \cite{globalanalysis}. Details of the underlying
formalism can again be found in \cite{DFHL,FHL,FLPR}. We work in the
limit where the quadratic solar mass splitting is ignored, i.e.
$\dm{21}=0$.  We have thus two leading oscillation parameters
($\theta_{23}$ and $\dm{}$) and two sub-leading parameters
($\theta_{13}$ and $\mathrm{sgn}\,\dm{}$) and no effects from the
CP-violating phase $\delta$ \cite{CP}. This approximation is justified
since CP-violating effects disappear quickly for the large distances
considered \cite{FLPR}.  For the leading oscillation parameters the
question is how precise they can be extracted and the goal for the
sub-leading parameter $\theta_{13}$ is to measure it with some
precision or to give at least an improved limit below the present
CHOOZ limit of $\sin^2 2\theta_{13}<0.1$ \cite{CHOOZ}. $\theta_{13}$
also dominantly controls the impact of matter effects in oscillation
measurements. Thus the $\theta_{13}$ reach is highly correlated with
the ability to determine the structure of the mass hierarchy (i.e the
sign of $\dm{}$).

\begin{figure}[ht!]
\begin{center}
\epsfig{file=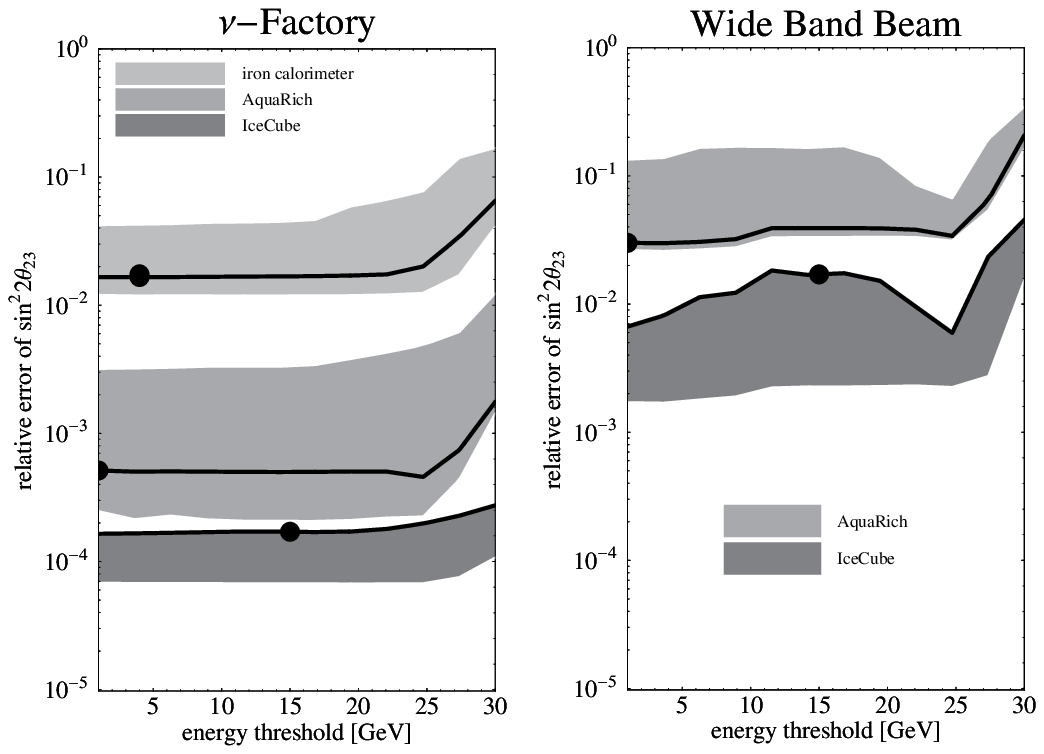,width=16cm}
\mycaption{Relative 3$\sigma$--errors of $\sin^22\theta_{23}$ 
as a function of the energy threshold of the detector at a baseline of 
11200~km. The colored bands show the influence of the energy resolution 
from 5\% (lower edge) up to 50\% (upper edge). The black lines indicate 
the energy resolution of a typical detector of the corresponding type
and the black dots mark the typical threshold value on this line 
as specified in table 1.}
\label{fig:theta23}
\end{center}
\end{figure}

\begin{figure}[htb!]
\begin{center}
\epsfig{file=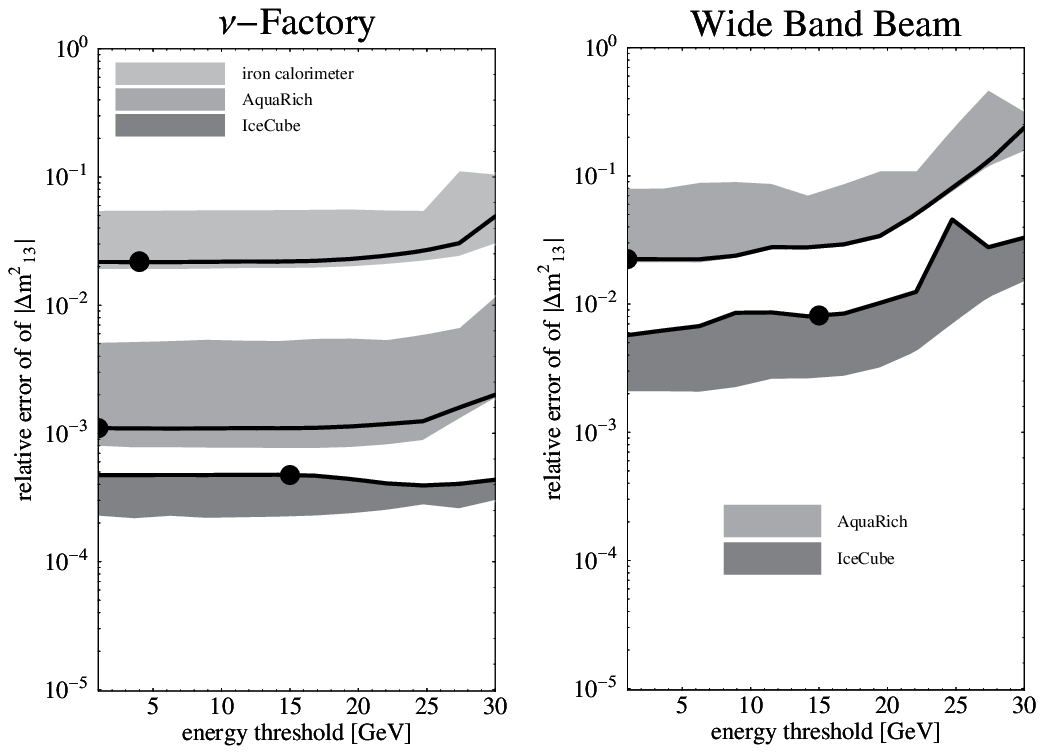 ,width=16cm}
\end{center}
\mycaption{Relative 3$\sigma$--error in $|\dm{31}|$
as a function of the energy threshold of the detector at a baseline of 11200~km. 
The colored bands show the influence of the energy resolution 
from 5\% (lower edge) up to 50\% (upper edge). The black lines indicate 
the energy resolution of a typical detector of the corresponding type
and the black dots mark the typical threshold value on this line 
as specified in table 1.}
\label{fig:dm31}
\end{figure}

\begin{figure}[htb!]
\begin{center}
\epsfig{file=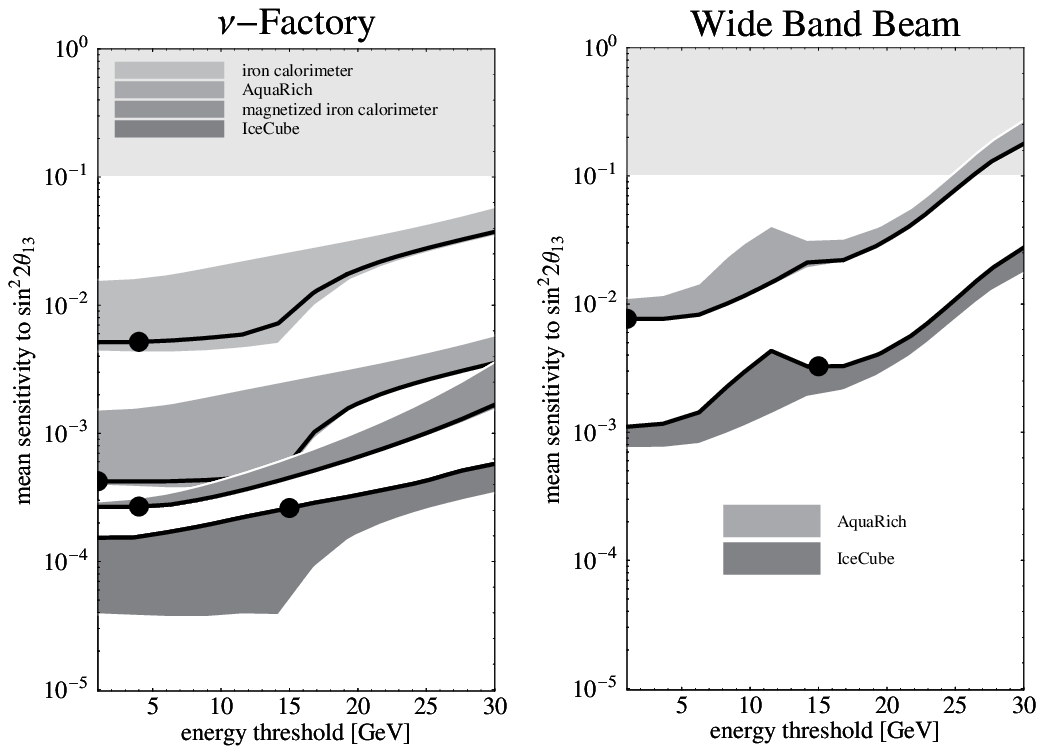,width=16cm}
\mycaption{Mean sensitivity at 90\% C.L. to $\sin^22\theta_{13}$
depending on the energy threshold of the detector at a baseline of 6500~km. 
The colored bands show the influence of the energy resolution 
from 5\% (lower edge) up to 50\% (upper edge). The black lines indicate 
the energy resolution of a typical detector of the corresponding type
and the black dots mark the typical threshold value on this line 
as specified in table 1. The light grey shaded area represents the CHOOZ limit.}
\label{fig:theta13}
\end{center}
\end{figure}

\begin{figure}[htb!]
\begin{center}
\epsfig{file=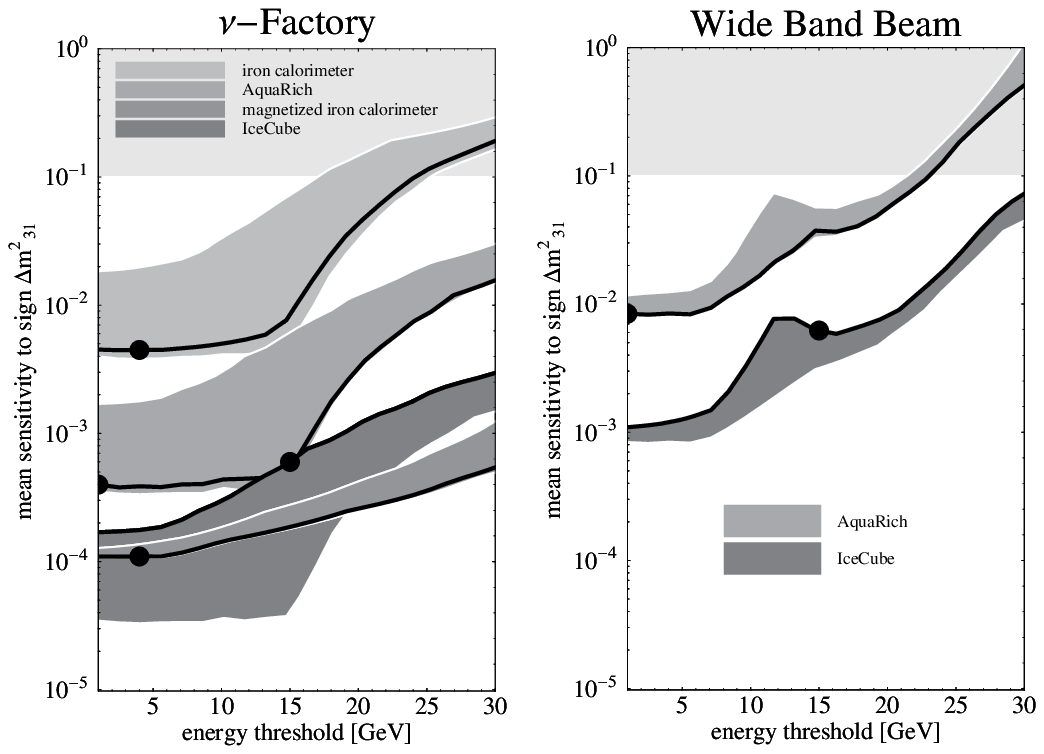,width=16cm}
\end{center}
\mycaption{Mean sensitivity in $\sin^22\theta_{13}$ at 90\% C.L. to the 
sign of $\dm{31}$
depending on energy threshold of the detector at a baseline of 6500~km. 
The colored bands show the influence of the energy resolution 
from 5\% (lower edge) up to 50\% (upper edge). The black lines indicate 
the energy resolution of a typical detector of the corresponding type
and the black dots mark the typical threshold value on this line 
as specified in table 1. The light grey shaded area represents the CHOOZ limit.}
\label{fig:sign}
\end{figure}

The following plots show the precision in the measurement of these
parameters for the three discussed detector types (magnetized iron
detector with and without CID, IceCube and AQUA-RICH) in combination
with the two beams described. We compare the performance to determine
the leading parameters $\theta_{23}$ (fig.~\ref{fig:theta23}) and
$|\dm{31}|$ (fig.~\ref{fig:dm31}).  Also shown is the sensitivity for
$\theta_{13}$ (fig.~\ref{fig:theta13}) and the capability to determine
the sign of $\dm{31}$ (fig.~\ref{fig:sign}). Fig.~\ref{fig:sign} shows
the limit in $\theta_{13}$ above which tests of matter effects are
possible.  A detailed explanation of how these results are obtained
can be found in \cite{FHL,FLPR} where the applied statistical method
is also presented.  The figures show always on the left side the
results for a neutrino factory beam, while the right plots show the
conventional CNGS--like Wide Band Beam.

The precisions and sensitivities in
the figures are presented as functions of the energy threshold ranging
from 1~GeV to 30~GeV. The dependence on the energy resolution is in
all cases strictly monotonic and nearly linear on logarithmic scales.
This dependence is shown in the plots by the bands for a resolution of
5\% (lower edge) to 50\% (upper edge). The black lines inside the
bands denote the characteristic energy resolution given above in table 1
 and the dot on these lines marks the
corresponding typical energy threshold. The baseline of 6500~km may
not be a realistic option for IceCube, but the influence of the
baseline for detectors without CID is between 6000~km and 12000~km
marginal. The sensitivity reach and the resolution scale in general
approximately as $\sqrt{M \Phi t}$, where $M$ is the detector mass,
$\Phi$ the neutrino flux and $t$ the running time of the experiment.

Figs.~\ref{fig:theta23} and \ref{fig:dm31} show the precision which
could in principle be obtained in the determination of 
$\sin^22\theta_{23}$ and $|\dm{31}|$ at a baseline of 11200~km. 
There is no difference in the performance of an 
iron detector with and without charge identification so that there
appears only one band for this type of experiment in the figure. 
This is not unexpected, since the information on these two parameters 
is contained in the disappearance rates. All experimental setups 
show not much influence of the energy threshold on the precision 
of the leading parameters. The accuracy is mainly governed by statistics 
which is given essentially by the mass of the detector. Thus IceCube, 
the largest detector in our study, has for both beams the best potential.

Fig.~\ref{fig:theta13} shows the mean sensitivity to $\sin^22\theta_{13}$. 
The magnetized iron detector benefits here significantly from charge
identification (CID), i.e. the capability to measure the appearance channel.
For the neutrino factory beam pointing to detectors without CID there is 
a strong threshold effect at about 15~GeV, which is roughly the 
MSW--resonance energy in the Earth mantle. This threshold effect is less 
pronounced for the Wide Band Beam since its flux is much lower in the
MSW--resonance energy regime. Fig.~\ref{fig:theta13} demonstrates nicely
the importance of the energy threshold for experiments which are based on 
conventional beams pointing to large Cherenkov detectors like IceCube or
ANTARES. Note that neutrino factory beams pointing to large water or ice 
Cherenkov detectors have in principle the highest potential if the
involved technological issues can be solved. A discussion of the
questions involved can be found in \cite{DFHL}. 
The strongest influence of the threshold is found for the determination 
of the the sign of $\dm{31}$ in fig.~\ref{fig:sign}. For the setups
with a neutrino factory beam there is again a clear effect at about 15~GeV, 
for the same reasons as mentioned above.

In summary, we have compared in this study the generic physics potential
for neutrino oscillation studies. We considered neutrino factory beams 
and conventional Wide Band Beams pointing to three different types 
of detectors: magnetized iron calorimeters, large water or ice Cherenkov 
detector and megaton ring imaging Cherenkov detectors. The quantities
studied are the precision for the leading oscillation parameters 
$\sin^22\theta_{23}$ and $|\dm{31}|$ as well as the sensitivity to the 
sub-leading MSW--enhanced oscillation parameters $\sin^22\theta_{13}$ 
and the sign of $\dm{31}$.
For the precision measurement of $\sin^22\theta_{23}$ and $|\dm{31}|$ 
the single most important parameter which characterizes the performance 
of a detector is its mass. Neither typical energy thresholds for muon 
detection nor the energy resolution have a big influence on the precision 
as long as the threshold does not exceed $30~\GeV$. Thus large km$^3$ 
Cherenkov detectors like IceCube in conjunction with a neutrino factory 
beam perform in principle best, with a relative error down to about 
$4\cdot10^{-4}$. A conventional beam pointed to an IceCube-like detector 
performs in principle about half an order of magnitude better than the 
usually considered combination of neutrino factory and magnetized iron 
calorimeter, which has an error of about $10^{-2}$. Note however that 
we discuss here the generic physics potential which can be achieved in
principle and further systematic and background effects may (and probably
will) affect our conclusions.

For the determination of the sub-leading parameters $\sin^22\theta_{13}$ 
and the sign of $\dm{31}$ we find a different situation in our results.
The possibility to measure appearance rates (requiring sufficient muon 
charge identification) is here extremely helpful. Especially for detectors 
without charge identification we observe an intricate interplay of event 
rates, threshold and energy resolution. To obtain useful results with a 
conventional beam one needs then very large detector masses of at least 
$1\thinspace\textrm{Mt}$. The threshold dependence of the limits is also 
rather strong and the energy resolution is also important. Huge IceCube-like 
detectors and the megaton ring imaging detector (AQUA-RICH) 
have however in principle a comparable performance down to 
$\sin^22\theta_{13}\simeq(2-8)\cdot10^{-3}$.
The wrong sign muon signal in a magnetized iron calorimeter gives for a 
neutrino factory beam the best sensitivity of about 
$\sin^22\theta_{13}\simeq10^{-4}$. The performance of IceCube is however
in principle rather similar and reaches also down to
$\sin^22\theta_{13}\simeq 4\cdot 10^{-4}$. 
Note however again that we study here the generic physics potential
and the measurement of $\sin^22\theta_{13}$ and the sign of $\dm{31}$ 
will be limited essentially in all cases by further systematic and 
background issues.

Our results show in a larger context that detectors with imperfect or 
even missing muon charge identification (which are therefore insensitive 
to the appearance oscillation channel) are useful for precision 
measurements of oscillation parameters \cite{DFHL}. Especially in 
high rate neutrino experiments, where statistical errors are small, 
the information which is inherent to the disappearance rates is very 
useful. 

The generic physics potential studied here is expected to be at least 
in some cases limited by further technological limitations, systematic 
and background effects. For magnetized iron detectors excellent charge 
separation is for example a crucial issue. Neutrino telescopes are 
in principle very interesting due to their huge mass. For the high
event rates obtained directional information becomes less important 
and it is possible to lower the threshold and to use them for 
oscillation physics. How well this works has to be further studied.
For Wide Band Beams the limiting factors could e.g. lie in the flux 
monitoring with a near detector. On the other side Wide Band Beam
technology will improve even further. 

Altogether Wide Band Beams have in principle a very good physics 
potential to improve our knowledge of $\sin^22\theta_{23}$, $|\dm{31}|$, 
$\sin^22\theta_{13}$ and the sign of $\dm{31}$ until a neutrino 
factory is built. It is thus conceivable that $\sin^22\theta_{13}$ 
and even the sign of $\dm{31}$ could be measured (or limited) with
conventional beams even down to $\sin^22\theta_{13}\simeq 10^{-3}$. 
A neutrino factory would enable measurements (or limits) of these 
parameters down to $\sin^22\theta_{13}\simeq 10^{-4}$ and in addition 
it would so far be the only realistic way to measure leptonic 
CP--violation \cite{CP}. 

%%%%%%%%%%%%%%%%%%%%%%%%%%%%%%%%%%%%%%%%%%%%%%%%%%%%%%%%%%%%%%%%%%%%%

 \newpage

\def\NL{\\[-1mm]}

\bibliographystyle{phaip}

%%%%%%%%%%%%%%%%%%%%%%%%%%%%%%%%%%%%%%%%%%%%%%%%%%%%%%%%%%%%%%%%%%%%%%%
\newpage

\ \vspace*{3.cm}

\end{document}